\documentclass[prl,preprint,showpacs,floatfix,superscriptaddress]{revtex4-1}

\usepackage{chemformula} % Formula subscripts using \ch{}
\usepackage[T1]{fontenc} % Use modern font encodings
\usepackage{dcolumn}

\usepackage{ulem}
\usepackage{makecell}
\usepackage[utf8]{inputenc} 
\usepackage{graphicx}
\usepackage{cleveref}
\usepackage{amsfonts}
\usepackage{amsmath,bm,amssymb}
\mathchardef\mhyphen="2D

\usepackage{kotex}
\usepackage{comment}
\usepackage{usebib} %related to citation

\begin{document}

\title{On the origin and the manipulation of ferromagnetism in Fe$_3$GeTe$_2$: defects and dopings}

\author{Seung Woo Jang}
\affiliation{Department of Physics, Korea Advanced Institute of Science and Technology (KAIST), Daejeon 34141, Republic of Korea}
\author{Min Yong Jeong}
\affiliation{Department of Physics, Korea Advanced Institute of Science and Technology (KAIST), Daejeon 34141, Republic of Korea}
\author{Hongkee Yoon}
\affiliation{Department of Physics, Korea Advanced Institute of Science and Technology (KAIST), Daejeon 34141, Republic of Korea}
\author{Siheon Ryee}
\affiliation{Department of Physics, Korea Advanced Institute of Science and Technology (KAIST), Daejeon 34141, Republic of Korea}
\author{Myung Joon Han}
\affiliation{Department of Physics, Korea Advanced Institute of Science and Technology (KAIST), Daejeon 34141, Republic of Korea}
\email{mj.han@kaist.ac.kr}

\begin{abstract}
	
To understand the magnetic properties of Fe$_3$GeTe$_2$, we performed the detailed first-principles study. Contrary to the conventional wisdom, it is unambiguously shown that Fe$_3$GeTe$_2$ is not ferromagnetic but antiferromagnetic carrying zero net moment in its stoichiometric phase. Fe defect and hole doping are the keys to make this material ferromagnetic, which are shown by the magnetic force response as well as the total energy calculation with the explicit Fe defects and the varied system charges. Further, we found that the electron doping also induces the antiferro- to ferromagnetic transition. It is a crucial factor to understand the notable recent experiment of gate-controlled ferromagnetism. Our results not only unveil the origin of ferromagnetism of this material but also show how it can be manipulated with defect and doping.	 

\end{abstract}

\maketitle


{\it Introduction --} Recently, magnetic 2-dimensional (2D) van der Waals (vdW) materials have attracted tremendous attention \cite{burch_magnetism_2018,gong_two-dimensional_2019,wang_raman_2016,tian_magneto-elastic_2016,lee_ising-type_2016,gong_discovery_2017,huang_layer-dependent_2017,bonilla_strong_2018,kim_charge-spin_2018,fei_two-dimensional_2018, zhong_van_2017,seyler_ligand-field_2018, jiang_electric-field_2018,jiang_controlling_2018,huang_electrical_2018,wang_electric-field_2018,song_giant_2018,klein_probing_2018,wang_very_2018,kim_one_2018,cardoso_van_2018,zhang_emergence_2018,deng_gate-tunable_2018,wang_tunneling_2018,li_patterning-induced_2018,kim_large_2018,tan_hard_2018,song_voltage_2019}. On the one hand, the magnetic order itself in 2D limit is an interesting physical phenomenon \cite{mermin_absence_1966}. From the point of view of device applications, on the other, they have great potentials for the electrical control of magnetism and spintronic device \cite{burch_magnetism_2018, gong_two-dimensional_2019, jiang_electric-field_2018, jiang_controlling_2018, huang_electrical_2018, wang_electric-field_2018, song_giant_2018, klein_probing_2018, song_voltage_2019}. Based on ferromagnetic (FM) materials, in particular, recent experiments have shown many  fascinating phenomena and useful possibilities such as the electronically tunable magnetism \cite{wang_electric-field_2018,deng_gate-tunable_2018}, room-temperature ferromagnetism \cite{bonilla_strong_2018, deng_gate-tunable_2018}, and the controlled spin and valley pseudospins \cite{zhong_van_2017}.

In this regard, a great amount of research efforts is now being devoted to  Fe$_3$GeTe$_3$ (FGT). Bulk FGT is known as a FM metal with a critical temperature of $T_c\simeq$ 220 K \cite{deiseroth_fe3gete2_2006,chen_magnetic_2013,verchenko_ferromagnetic_2015,may_magnetic_2016}. Its 2D form has recently been reported by Fei {\it et al} who showed that the ferromagnetism of this material survives down to mono-layer \cite{fei_two-dimensional_2018}. Furthermore, Deng {\it et al} demonstrated that $T_c$ can be controlled by gating, and can eventually reach to room temperature \cite{deng_gate-tunable_2018}. It is certainly a useful feature for device applications. Other intriguing and promising aspects of FGT include the tunneling spin valve \cite{wang_raman_2016}, patterning-induced ferromagnetism \cite{li_patterning-induced_2018}, thickness-dependent hard magnetic phase \cite{tan_hard_2018}, large anomalous Hall current driven by topological nodal lines \cite{kim_large_2018}, and Kondo behavior \cite{zhang_emergence_2018}.

In this Letter, we start with a motivation that the origin of ferromagnetism of FGT is still not clearly understood \cite{yi_competing_2017}. First of all, we show that the magnetic ground state of stoichiometric FGT is not FM, which is in sharp contrast to the various experimental reports. Our extensive calculations of total energies and magnetic force responses establish that the inter-layer coupling of FGT is antiferromagnetic (AFM). Second, it is defects that make this material FM. It is found that introducing a small amount of Fe defects or hole doping  quickly changes the inter-layer coupling to be FM. This conclusion is consistent with the previous experimental reports about the difficulty in synthesizing stoichiometric samples without Fe deficiency.  Finally, we demonstrate that FM order can also be induced by electron doping.  It is of crucial importance to understand the recent experiment by Deng {\it et al} \cite{deng_gate-tunable_2018}. Our current work sheds new light on understanding the origin of ferromagnetism in FGT and how to manipulate it through defect and doping.

%\section{Result and Discussion}

\begin{figure}
	\centering
	\includegraphics[width=0.8\linewidth]{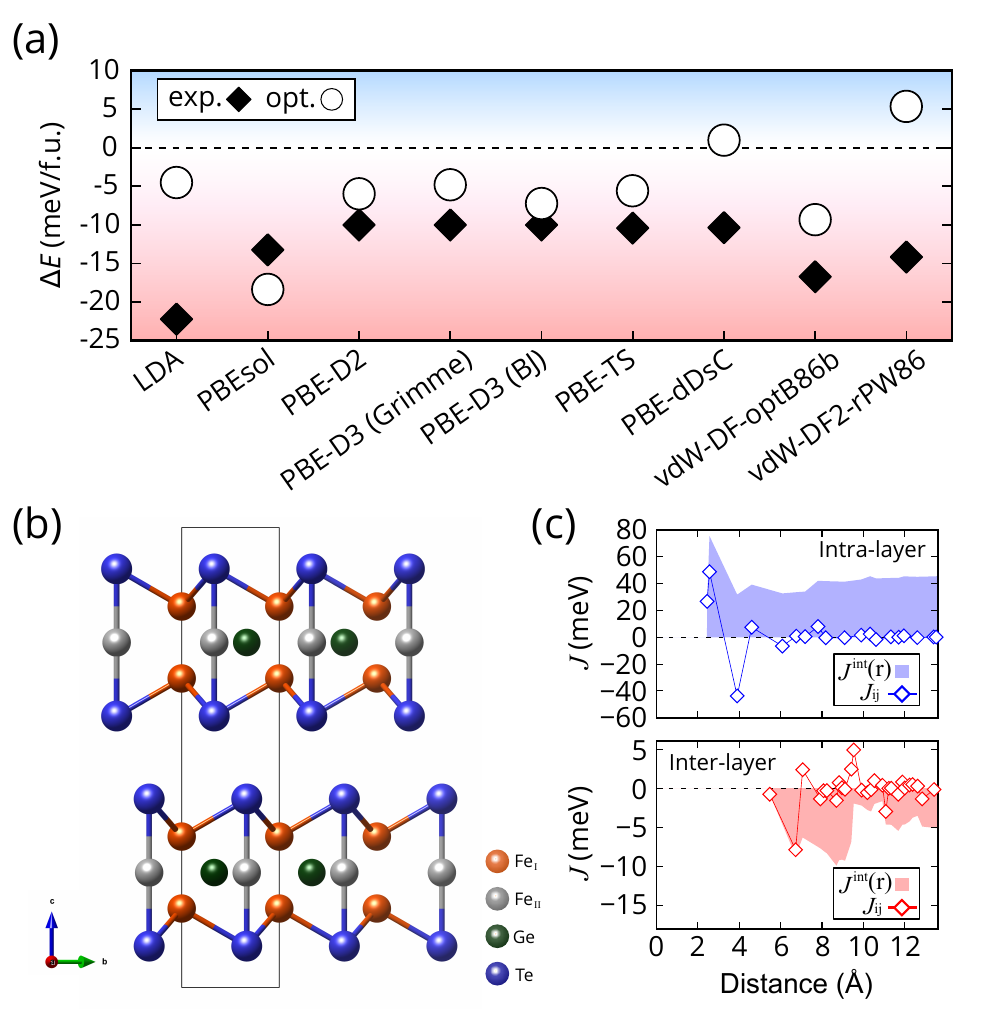}
	\caption{The magnetic interaction and the ground state of undoped FGT. (a) The calculated total energy differences ($\Delta E$) between AFM and FM inter-layer order using various XC-vdW functionals. The results from the experimental \cite{deiseroth_fe3gete2_2006} and the optimized structures are indicated by diamonds (black) and circles (white), respectively. (b) The side view of FGT structure. Fe$_{\textrm{I}}$, Fe$_{\textrm{II}}$, Ge, and Te atoms are represented by orange, gray, green, and blue spheres, respectively. Fe$_{\textrm{I}}$ and Fe$_{\textrm{II}}$ represent two inequivalent Fe sites with $+3$ and $+2$ formal charge state, respectively. (c) The calculated magnetic couplings, $J_{ij}$, as a function of inter-atomic pair distance (also taking the coordination number into account; {\it i.e.}, all equal distance interaction represented by one symbol). The intra- and inter-layer interactions are represented in the upper and lower panel, respectively. The filled curves (shaded areas) show the integrated values of $J_{ij}$ up to the given distance; $J^{\rm int}(r)=\sum^{r}_{0} J_{ij}$.}
	\label{fig:fig1}
\end{figure}

{\it Result(1): AFM ground state -- }Figure 1(a) presents the calculated total energy difference between the inter-layer (out-of-plane) AFM and FM spin order; $\Delta E = E_{\rm AFM} - E_{\rm FM} $ (within the plane, FM order is always favored; not shown). We pay special attention to the fact that describing vdW interaction within density functional theory (DFT) framework needs care because the exchange-correlation (XC) functional form for this weak interaction has not yet been quite well established \cite{grimme_semiempirical_2006,grimme_consistent_2010,dion_van_2004,roman-perez_efficient_2009,grimme_effect_2011,klimes_van_2011}. As a practical way to resolve this issue and to investigate the ground state spin order, we adopted 9 different functional forms \cite{ceperley_ground_1980,perdew_self-interaction_1981,PBE,perdew_restoring_2008,grimme_semiempirical_2006,grimme_consistent_2010,grimme_effect_2011,tkatchenko_accurate_2009,steinmann_comprehensive_2011,steinmann_generalized-gradient_2011,dion_van_2004,roman-perez_efficient_2009,klimes_van_2011,lee_higher-accuracy_2010}. We also considered both the experimental (black diamonds) and the optimized structure (white circles). It is clearly seen that the AFM inter-layer coupling is energetically more stable in most of the functionals. The only exception is the fully-relaxed structure with so-called `vdW-DF2-rPW86' functional \cite{lee_higher-accuracy_2010}. This functional, however, significantly overestimates the lattice parameters (see Supplemental Material Sec.~1 for more details). For the case of `PBE-dDsC' functional \cite{steinmann_comprehensive_2011,steinmann_generalized-gradient_2011}, FM order is slightly more favorable, but its energy difference from AFM is too small ($\sim$ 0.98 meV/f.u.). We hereby conclude that the inter-layer magnetic configuration of FGT is AFM.

In order to nail down this conclusion, namely, the AFM inter-layer coupling, we performed magnetic force theory (MFT) calculation which measures the magnetic moment response at one atomic site to the perturbative change of another \cite{MFT1,MFT2,han_electronic_2004,yoon_reliability_2018}. The results are summarized in Figure 1(c); the intra-layer (within the plane) and inter-layer magnetic couplings are presented in the upper and lower panel, respectively. Since our MFT calculation is conducted in the momentum space and then transformed into the real space \cite{yoon_reliability_2018}, all inter-atomic pair interactions are obtained in a single response computation. In Figure 1(c), we present them as a function of inter-atomic distance by taking into account the coordination number. As the inter-atomic distance becomes larger, the magnetic interaction strength $J_{ij}$ gets reduced and eventually becomes zero as expected. The first thing to be noted is the FM intra-layer coupling. The integrated magnetic interaction $J^{\rm int}(r)$ ({\it i.e.,} the sum of all pair interactions $J_{ij}$ up to the given distance $r$; represented as the colored area) clearly shows that the intra-layer magnetic interaction is always FM, and it is dominated by the first two FM couplings.

Importantly, the inter-layer magnetic interaction is AFM. The lower panel of Figure~1(c) shows that, although there are many FM inter-atomic pair interactions along the out-of-plane direction, the overall interaction is steadily AFM; $J^{\rm int}(r)$ remains negative (the shaded area). The second neighbor AFM pair dominates the inter-layer interaction and stabilizes AFM order. We emphasize both total energy and magnetic force calculation unequivocally declare that the stoichiometric FGT is AFM carrying zero net moment.

\begin{figure}
	\centering
	\includegraphics[width=0.8\linewidth]{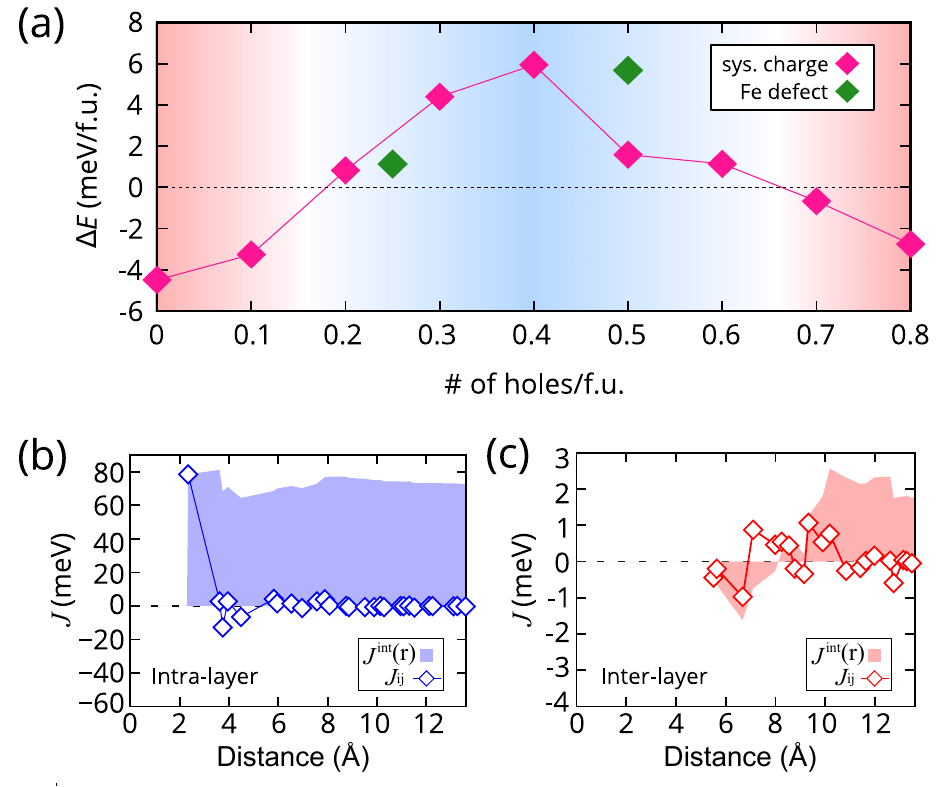}
	\caption{The magnetic interaction and the ground state of hole-doped FGT. (a) The calculated total energy differences ($\Delta E$) between the inter-layer AFM and FM phases as a function of hole concentration. The magenta and green symbols are the results of the calculation with varying system charge and the supercell calculation with Fe defects, respectively. For the latter, the doping concentration is determined by assigning three and two holes for one deficient Fe$_{\textrm{I}}$ and Fe$_{\textrm{II}}$, respectively. (b, c) The calculated intra-layer (b) and inter-layer (c) magnetic couplings obtained from the supercell calculation of Fe$_{2.75}$GeTe$_2$ (0.5 hole doping per f.u.). 
	The calculated $J_{ij}$ and $J^{\rm int}(r)$ are represented with diamond symbols and the filled curves (shaded areas), respectively.}
	\label{fig:fig2}
\end{figure}

{\it Result(2): Fe defect, hole-doping and FM ground state -- }Seemingly, this conclusion is  in a sharp contradiction to the various experiments because FGT is known as a FM metal for both bulk and thin film. Here we identify, however, that the crucial factor to stabilize ferromagnetism is the hole doping. According to the literature, it seems quite difficult to make defect-free stoichiometric FGT, and the sample gets easily Fe-deficient \cite{deiseroth_fe3gete2_2006,verchenko_ferromagnetic_2015,may_magnetic_2016,zhu_electronic_2016,liu_critical_2017-1,liu_anomalous_2018}. Since the Fe deficiency induces the hole doping, we investigate its effect on the magnetic order which has never been studied before.

First, we calculated the total energy difference $\Delta E$ as a function of doping. Figure 2(a) clearly shows that introducing $\sim$0.2--0.6 holes per formula unit (f.u.) induces the AFM to FM transition. This level of hole doping is well consistent with the Fe defect concentration reported in experiments: $0.11 < x < 0.36$ for single crystalline Fe$_{3-x}$GeTe$_2$ or powder, \cite{deiseroth_fe3gete2_2006,verchenko_ferromagnetic_2015,may_magnetic_2016,zhu_electronic_2016,liu_critical_2017-1,liu_anomalous_2018} and $0.03 < x < 0.31$ for polycrystalline sample\cite{may_magnetic_2016} although its relationship to the magnetic ground state has never been discussed nor speculated. We think that a similar amount of Fe defects can likely be present also in the thin-film samples.

Second, we perform the supercell total energy calculation to directly simulate the Fe-deficient sample. In our supercell setups, Fe$_{2.75}$GeTe$_2$ and Fe$_{2.875}$GeTe$_2$ can be simulated, which are well compared with experiments \cite{deiseroth_fe3gete2_2006,verchenko_ferromagnetic_2015,may_magnetic_2016,zhu_electronic_2016,liu_critical_2017-1,liu_anomalous_2018}. The results are presented with green diamond symbols in Figure~2(a). It clearly shows that, with Fe defects, the FM order is energetically favorable, $\Delta E >0$.

Finally, we perform MFT calculation to double check this conclusion. The results of Fe$_{2.75}$GeTe$_2$ are summarized in Figure~2(b) and (c) which presents the intra- and inter-layer interaction, respectively. Note that the inter-layer interaction changes to be FM ({\it i.e.}, the shaded area of $J_{\rm int}$ becomes positive in Figure~2(c)) while the intra-layer coupling remains FM (see Figure~2(b)). Once again, this is a strong and independent confirmation that, upon hole doping, the inter-layer magnetic order is changed to FM. (For further evidences that the defect and hole doping critically affect the magnetic property of FGT, see Supplemental Material Sec.~2 and 3.)

\begin{figure}
	\centering
	\includegraphics[width=0.8\linewidth]{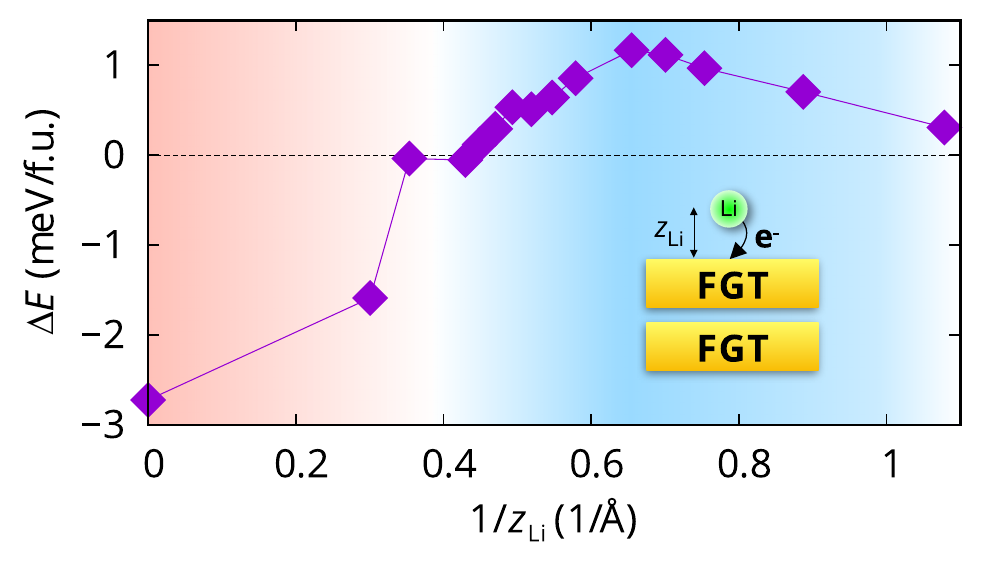}
	\caption{The calculated $\Delta E$ as a function of the inverse height ($1/z_{\rm Li}$) of Li atom. $1/z_{\rm Li}$ is proportional to the amount of doped electrons in FGT layers. The inset shows the schematic illustration of the electron doping from Li onto the FGT.}
	\label{fig:figure-3}
\end{figure}

{\it Result(3): Electron doping and the recent gating experiment -- }Once we establish  that stoichiometric FGT is AFM and that FM order is induced by Fe defects or hole dopings, an important new question arises. Recently, Deng {\it et al} reported the gate tunability of ferromagnetism in ultra-thin FGT \cite{deng_gate-tunable_2018}. It is demonstrated that the magnetic property including $T_c$ can be enhanced and manipulated by ionic gating. Here we note that their gating dopes electrons, not holes.

In order to understand the experiment and to answer this fundamentally important question, we investigate the electron doping effect on the magnetic order. Remarkably, we found, the electron doping also induces FM order. To simulate the experimental situation \cite{deng_gate-tunable_2018}, Li atom is placed in the vicinity of the top surface of FGT bi-layer; see the inset of Figure~3. Due to the high electropositivity of Li, electrons are transferred and doped into FGT. Figure~3 presents the calculated $\Delta E$ as a function of the inverse height of Li atom, $1/z_{\rm Li}$  ($z_{\rm Li}$ is defined as the distance between Li and the top surface of FGT). $1/z_{\rm Li}$ was reported to well represent the amount of doped electrons \cite{kim_observation_2015,baik_emergence_2015} (for more details, see Supplemental Material Sec.~4). Note that the AFM to FM transition is induced by electron doping: In the high doping regime ($0.5 \leq 1/z_{\rm Li} \leq 1$), the calculated $\Delta E$ is positive. As the height increases ({\it i.e.}, $1/z_{\rm Li} \longrightarrow 0$), FGT eventually becomes AFM as expected, corresponding to the undoped FGT. This result elucidates the relation between the electron doping and the ferromagnetism of FGT, thereby establishing the physical picture for the recent experiment by Deng {\it et al} \cite{deng_gate-tunable_2018}.

{\it Discussion -- }Our result provides further information to understand the more details of the gating experiment. An intriguing feature observed in Ref.~\onlinecite{deng_gate-tunable_2018} is that the measured $T_c$ and the coercive field exhibit a clear deep at around  1.4 volt before the rapid rise at around 1.8 volt which indicates that ferromagnetism is first suppressed and then revived as a function of electron doping. Our current study suggests that this behavior can at least partly be related to the interplay between the Fe defects and the doped electrons. Also, from the fact that the defect-free FGT is antiferromagnetically ordered along the out-of-plane direction, one can expect to realize the CrI$_3$-type magnetic geometry from this metallic material, which was suggested to be useful for device applications \cite{huang_layer-dependent_2017,seyler_ligand-field_2018,jiang_electric-field_2018,jiang_controlling_2018,huang_electrical_2018,klein_probing_2018,song_giant_2018}.

{\it Summary and conclusion -- }To summarize, we, for the first time, identify the origin of ferromagnetism in FGT. Our total energy and MFT calculation unambiguously show that the defect-free stoichiometric FGT is AFM while introducing Fe defects or holes stabilizes the FM solution. Further, it is demonstrated that the electron doping induces the AFM to FM transition as well. It is crucially important to understand the recent gate control experiment in which the ionic gating dopes the thin film FGT with electrons. \\[3mm]

{\it Computational methods --} We performed total energy and electronic structure calculations employing DFT as implemented in VASP (Vienna Ab initio Software Package) \cite{vasp}. The local density approximation (LDA) XC functional \cite{ceperley_ground_1980} as parameterized by Perdew and Zunger \cite{perdew_self-interaction_1981} was used unless specified otherwise. To confirm the robustness of our conclusion, we double checked the undoped case with many different XC functionals and vdW corrections including PBE \cite{PBE}, PBEsol \cite{perdew_restoring_2008}, D2 \cite{grimme_semiempirical_2006}, D3 (Grimme) \cite{grimme_consistent_2010,grimme_effect_2011}, D3 (BJ) \cite{grimme_consistent_2010,grimme_effect_2011}, TS \cite{tkatchenko_accurate_2009}, dDsC \cite{steinmann_comprehensive_2011,steinmann_generalized-gradient_2011}, vdW-DF-optB66b \cite{dion_van_2004,roman-perez_efficient_2009,klimes_van_2011}, and vdW-DF2-rPW86 \cite{lee_higher-accuracy_2010}. For the bulk FGT, 700~eV energy cutoff and 18 $\times$ 18 $\times$ 4 {\bf k} mesh in the first Brillouin zone were used. It is important to use large enough {\bf k} point grid (for the related discussion, see Supplemental Material Sec.~5). The force criterion for the structure optimization was 5 meV/{\AA}. For simulating the hole doping, we performed both calculations with the varying system charge and the explicit Fe defect. For the latter, two different supercells were considered, corresponding to the four- and eight-times of f.u. with a single Fe vacancy. The defect is created on the Fe$_{\textrm{II}}$ site as known from experiment \cite{may_magnetic_2016}. In order to simulate the gating experiment of electron dopings \cite{deng_gate-tunable_2018}, we performed the slab calculations with Li atom on top of bi-layer FGT. For bi-layer structures, the vacuum distance of $\sim$ $25$~{\AA} was used with 600~eV energy cutoff and 14 $\times$ 14 $\times$ 1 {\bf k} points. The amount of electron doping was controlled by adjusting the Li-atom positions. For MFT calculations \cite{MFT1,MFT2,han_electronic_2004,yoon_reliability_2018}, we used our DFT code, OpenMX \cite{openmx,LCPAO}.

({\it Additional note}) Throughout the manuscript, we assumed that the previous studies all agree that FGT is a FM metal. But there is a notable exception. Ref.\onlinecite{yi_competing_2017} reports the possible AFM coupling along the out-plane direction based on the magnetic measurement and DFT calculation. It is not clear how widely this claim is accepted in the later experimental studies \cite{tan_hard_2018,wang_tunneling_2018,li_patterning-induced_2018,fei_two-dimensional_2018,deng_gate-tunable_2018}. At least, however, their DFT calculation is certainly valid and indeed consistent with ours.

{\it Acknowledgments -- }This work was supported by Basic Science Research Program through the National Research Foundation of Korea (NRF) funded by the Ministry of Education (2018R1A2B2005204) and Creative Materials Discovery Program through the NRF funded by Ministry of Science and ICT (2018M3D1A1058754).

\end{document}

\title{Supplemental Material for ``On the origin and the manipulation of ferromagnetism in Fe$_3$GeTe$_2$: defects and dopings''}

	\author{Seung Woo Jang}
	\affiliation{Department of Physics, Korea Advanced Institute of Science and Technology (KAIST), Daejeon 34141, Republic of Korea}
	\author{Min Yong Jeong}
	\affiliation{Department of Physics, Korea Advanced Institute of Science and Technology (KAIST), Daejeon 34141, Republic of Korea}
	\author{Hongkee Yoon}
	\affiliation{Department of Physics, Korea Advanced Institute of Science and Technology (KAIST), Daejeon 34141, Republic of Korea}
	\author{Siheon Ryee}
	\affiliation{Department of Physics, Korea Advanced Institute of Science and Technology (KAIST), Daejeon 34141, Republic of Korea}
	\author{Myung Joon Han}
	\affiliation{Department of Physics, Korea Advanced Institute of Science and Technology (KAIST), Daejeon 34141, Republic of Korea}
	\email{mj.han@kaist.ac.kr}

\maketitle

\section{1. The lattice parameters and `\MakeLowercase{r}PW86' functional}

Table~\ref{Table S1} shows the optimized lattice parameters for the undoped bulk FGT obtained by various XC functionals. The experimental data is the value influenced by Fe defects \cite{deiseroth_fe3gete2_2006,may_magnetic_2016}. We note that `rPW86' functional significantly overestimates both $a$ and $c$.\\[4mm]

\begin{table}[h]%[!b]
	\begin{tabular}{lcccccccccc}
		\hline
		& \multicolumn{9}{c}{Optimized in FM} \\
		& LDA & PBEsol & D2 & D3(Grimme) & D3(BJ) & TS & dDsC & optB86b & rPW86  & Exp.\cite{deiseroth_fe3gete2_2006} \\
		\hline
		$a$ (\AA) &   3.90  & 3.98 & 4.00 & 4.02 & 4.01 & 4.01 & 4.02 & 4.02 &  4.17 & 3.99 \\ 
		$c$ (\AA) &   15.87 & 16.25 & 16.48  & 16.18 & 15.98 & 15.75 & 16.28 & 16.38 & 17.46 & 16.33 \\ 
		\hline
	\end{tabular}
	\caption{The structural parameters of FGT optimized by 9 different XC functionals.}
\label{Table S1}
\end{table}

\newpage
\section{2. FM stabilization energy and $T_c$ trend}

\begin{figure}[h]
	\begin{center}
		\includegraphics[width=0.65\linewidth,angle=0]{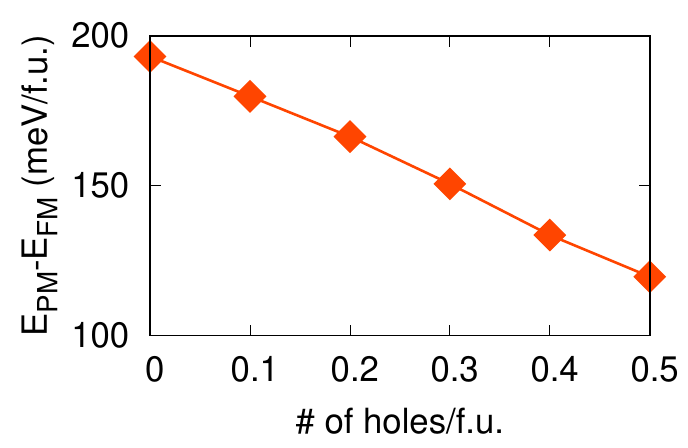}
		\caption{The calculated total energy difference between FM and PM solutions as a function of hole doping.
			\label{Figure S4}}
	\end{center}
\end{figure}

Figure~\ref{Figure S4} shows another meaningful indication that Fe defect and hole doping are of key importance to understand the ferromagnetism in FGT. The calculated total energy difference between FM and paramagnetic (PM) phase is presented as a function of hole doping. Note that this energy difference can be a measure of FM $T_c$. It monotonically decreases as the hole doping increases, which is in good agreement with the $T_c$ trend observed in experiment \cite{may_magnetic_2016}. Therefore, this result provides a further evidence that the ferromagnetism of FGT is governed by Fe defect and hole doping.\\[5mm]

\newpage
\section{3. The calculated magnetic moment for hole-doped FGT}

\begin{figure}[h]
	\begin{center}
		\includegraphics[width=0.65\linewidth,angle=0]{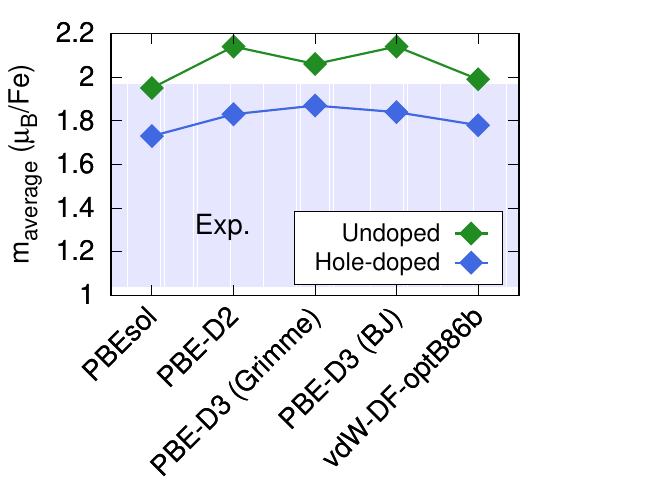}
		\caption{The calculated magnetic moment of bulk FGT by GGA-based XC functionals. The green and blue symbols represent the results of undoped and hole-doped FGT, respectively. We considered 0.5 holes per f.u., which corresponds to the FM region of phase diagram. The blue-shaded area shows the region of experimental values. \cite{deiseroth_fe3gete2_2006,chen_magnetic_2013,may_magnetic_2016,zhu_electronic_2016}. 
			\label{Figure S3}}
	\end{center}
\end{figure}

The calculated magnetic moment is another important indication of that the residing holes or Fe defects affect the magnetic property of FGT. 
Figure~\ref{Figure S3} presents the calculated Fe moment (the averaged value of Fe$_{\rm I}$ and Fe$_{\rm II}$) by several different GGA-based XC functionals. It has been reported that the use of GGA functional significantly overestimates the magnetic moment \cite{zhu_electronic_2016,zhuang_strong_2016}. It leads researchers to adopt LDA (plus vdW correction) which gives the better agreement with experiments. Figure~\ref{Figure S3} clearly shows that, by taking into account of hole dopings, the moment is noticeably reduced, and importantly, the results of all XC functionals fall into the range of experiment \cite{deiseroth_fe3gete2_2006,chen_magnetic_2013,may_magnetic_2016,zhu_electronic_2016}.\\[3mm]

\newpage
\section{4. $z_{\rm Li}$ and the electron doping}
\begin{figure}[h]
	\begin{center}
		\includegraphics[width=0.65\linewidth,angle=0]{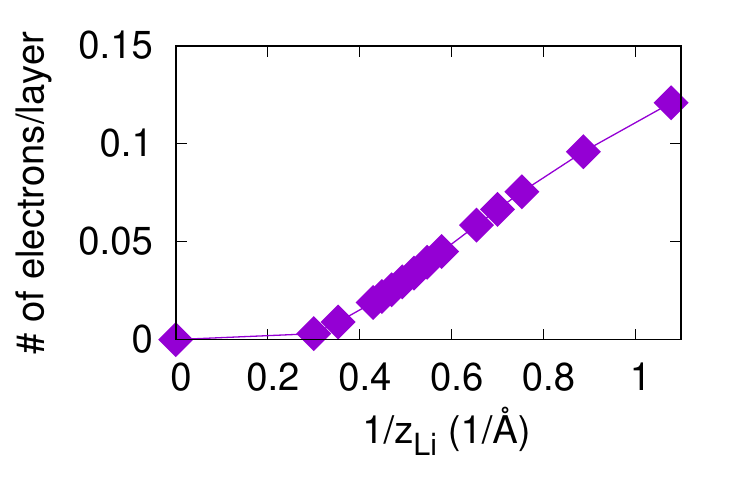}
		\caption{The number of doped electrons (Mulliken charge) as a function of the inverse distance between surface Te and Li atom.
			\label{Figure S5}}
	\end{center}
\end{figure}

In the main manuscript, we present the effect of electron doping in terms of Li height, $1/z_{\rm Li}$. Of course, it can also be presented as a function of Mulliken charge. Figure~\ref{Figure S5} shows the amount of doped or transferred electrons from Li to the bi-layer FGT. It is  proportional to the inverse height of Li atom.\\[5mm]

\newpage

\section{5. The \MakeLowercase{k}-mesh dependence of total energy}

\begin{figure}[h]
	\begin{center}
		\includegraphics[width=0.65\linewidth,angle=0]{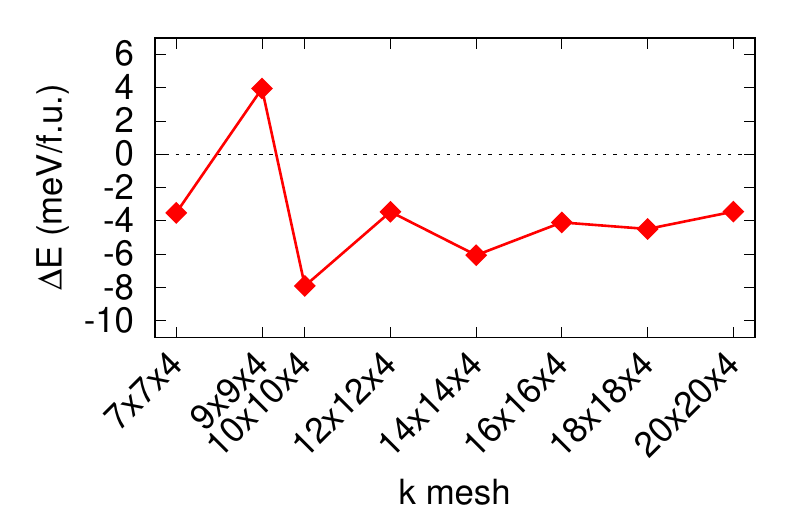}
		\caption{The calculated total energy difference ($\Delta E=E_{\rm AFM}-E_{\rm FM}$) for the undoped FGT with different {\bf k} meshes.
			\label{Figure S1}}
	\end{center}
\end{figure}

It is important to adopt a large enough number of {\bf k} points. As Figure~\ref{Figure S1} shows, 9$\times$9$\times$4 mesh is not enough which incorrectly gives the FM solution as the ground state. Using the denser {\bf k} meshes gives rise to the AFM ground state.

\providecommand{\latin}[1]{#1}
\makeatletter
\providecommand{\doi}
{\begingroup\let\do\@makeother\dospecials
	\catcode`\{=1 \catcode`\}=2 \doi@aux}
\providecommand{\doi@aux}[1]{\endgroup\texttt{#1}}
\makeatother
\providecommand*\mcitethebibliography{\thebibliography}
\csname @ifundefined\endcsname{endmcitethebibliography}
{\let\endmcitethebibliography\endthebibliography}{}